\documentclass[aps,amsmath,amsfonts]{revtex4}
\usepackage{feynmf}

\DeclareMathOperator{\tr}{Tr}

\newcommand{\nn}{\nonumber} 
\newcommand{\eq}[1]{eq.~(\ref{#1})}

\newcommand{\sla}[1]{#1 \! \! \! \slash}

\renewcommand{\bar}[1]{\overline{#1}}

\begin{document}

\preprint{ \hbox{MIT-CTP-3507} 
           \hbox{hep-ph/ 0406233} }

\title{Pentaquark masses in chiral perturbation theory}

\author{Vivek Mohta}
\affiliation{Center for Theoretical Physics and \\ Laboratory for Nuclear Science, \\ Massachusetts Institute of Technology \\ Cambridge, MA 02139\\ \\ Department of Mathematics, \\ Harvard University \\ Cambridge, MA 02138}

\begin{abstract}
Heavy baryon chiral perturbation theory for pentaquarks is applied beyond leading order. The mass splitting in the pentaquark anti-decuplet is calculated up to NNLO. An expansion in the coupling of pentaquarks to non-exotic baryons simplifies calculations and makes the pentaquark masses insensitive to the pentaquark-nucleon mass difference. The possibility of determining coupling constants in the chiral Lagrangian on the lattice is discussed. Both positive and negative parities are considered.
\end{abstract}

\maketitle

\section{Introduction} \label{intro}
Recent experiments have found evidence for an exotic hadronic resonance, the $\Theta^+(1540)$ \cite{Nakano:2003qx, Barmin:2003vv, Stepanyan:2003qr, Barth:ja, Asratyan:2003cb, Alt:2003vb, Kubarovsky:2003fi, Airapetian:2003ri,  Aleev:2004sa, Chekanov:2004kn}. This has initiated an effort in identifying other members of the same multiplet and answering a host of questions about these resonances. What are their masses and widths? Why are their widths so small?  What is their spin and parity? How are they produced? How certain is their existence?

Various models have been proposed for the pentaquarks that answer these questions with varying degrees of success. It would also be helpful to have a set of model independent results using a minimal set of assumptions common to the models. This suggests developing an effective field theory for pentaquarks. Since processes of interest, such as pentaquark decays, produce nucleons and soft mesons, it is natural to consider chiral perturbation theory including these fields.

In \cite{Ko:2003xx}, chiral perturbation theory is applied at leading order to decays of pentaquarks. However, in the formulation of the theory used, baryon masses in the Lagrangian make the power counting beyond leading order difficult. With some tweaking, namely the treatment of baryons as heavy fermions, the chiral perturbation theory offers a simple power counting and ease of calculation of Feynman diagrams \cite{Jenkins:1990jv}. In this form, it has been applied at leading order to decays of pentaquarks of both positive and negative parity and of both spin $\frac{1}{2}$ and $\frac{3}{2}$ \cite{Mehen:2004dy}. The results for positive parity and spin $\frac{1}{2}$ agree with those in \cite{Ko:2003xx} up to higher order corrections, as expected.

In the past, heavy baryon chiral perturbation theory has been applied to the octet and decuplet of baryons and to heavy mesons. The theory has been used to systematically calculate next-to-leading order (NLO) corrections in nucleon-nucleon scattering that agree with experiment \cite{Ecker:1994gg, Casalbuoni:1996pg}. The theory has also been used to calculate corrections to the Gell-Mann-Okubo formula for baryon masses \cite{Jenkins:1991ts}.The results have been useful in the chiral extrapolation of lattice predictions for baryon masses \cite{Leinweber:1999ig}.

In this paper, we apply heavy baryon chiral perturbation theory to pentaquarks beyond leading order. We calculate the mass splitting in the pentaquark anti-decuplet up to NNLO. We also identify an expansion in the coupling of pentaquarks to non-exotic baryons that simplifies calculations.

Throughout, we assume that $\Theta^+(1540)$ is a member of an anti-decuplet. Experiments suggest that $\Theta^+(1540)$ is an iso-singlet with hypercharge two. The smallest flavor multiplet that can accommodate it is an anti-decuplet. Such a multiplet of pentaquarks appears naturally in the diquark model, along with a degenerate octet of pentaquarks \cite{Jaffe:2003sg}. An anti-decuplet also appears in chiral soliton models \cite{Diakonov:1997mm} and in uncorrelated quark models, along with several other multiplets of pentaquarks of different masses. In this paper, we ignore the possibility of other pentaquark multiplets with a similar mass. We will take up the consequences of including other pentaquark multiplets in a later work. For most of the paper we assume that the pentaquark fields have spin one-half and positive parity. However, we also state the results for negative parity to show that the mass splitting does not distinguish between parities.

In section \ref{CL}, we review heavy baryon chiral perturbation theory for pentaquarks and develop a small coupling expansion. In section \ref{count}, we develop the power counting for this theory and determine the diagrams contributing to the self-energy at NNLO. In section \ref{mass}, we calculate the mass splitting in the pentaquark anti-decuplet at NNLO. In section \ref{negative} we treat negative parity. The final section has conclusions and directions for future work.

\section{Chiral Lagrangian and Small Coupling Expansion} \label{CL}
    The usual chiral perturbation theory for the pseudo-Goldstone bosons can be supplemented by the CCZW method to incorporate other ``matter fields" into the theory \cite{Coleman:sm, Callan:sn}. The method prescribes the most general Lagrangian for matter fields, that is invariant under chiral symmetry. We briefly review our field definitions and conventions and then construct the desired Lagrangian.  The pseudo-Goldstone bosons are described by the field $\xi = \exp (i \pi /f)$ where $f = 131 \, \textrm{MeV}$ is the pion decay constant and the field $\pi$ is 
\begin{equation}\label{pifield}
\pi = \begin{bmatrix}
\frac{1}{\sqrt{2}}\pi^0 + \frac{1}{\sqrt{6}}\eta & \pi^+ & K^+ \\
\pi^- & -\frac{1}{\sqrt{2}}\pi^0 +\frac{1}{\sqrt{6}}\eta & K^0 \\
K^- & \bar{K}{}^0 & -\frac{2}{\sqrt{6}}\eta \\
\end{bmatrix}.
\end{equation}
Under a chiral symmetry transformation, the field $\xi$ transforms as
\begin{equation}\label{xitrans}
\xi(x) \rightarrow L\ \xi(x)\ U^\dagger(x) = U(x)\ \xi(x)\ R^\dagger,
\end{equation}
where $U(x)$ defined in terms of $L$, $R$, and $\xi(x)$ by \eq{xitrans}.

The octet of baryon fields, which we refer to as nucleons from now on, is
\begin{equation}\label{bfield}
B = \begin{bmatrix}
\frac{1}{\sqrt{2}}\Sigma^0 + \frac{1}{\sqrt{6}}\Lambda & \Sigma^+ & p \\
\Sigma^- & -\frac{1}{\sqrt{2}}\Sigma^0 + \frac{1}{\sqrt{6}}\Lambda & n \\
\Xi^- & \Xi^0 & \frac{2}{\sqrt{6}}\Lambda\\
\end{bmatrix}.
\end{equation}
Under a chiral symmetry transformation, the field $B$ transforms as
\begin{equation}\label{btrans}
B(x)\rightarrow U(x) B(x) U(x)^\dagger
\end{equation}
where $U(x)$ is the transformation that appeared in \eq{xitrans}.
In fact, the CCZW results show that any transformation for $B$ that reduces to the adjoint transformation for the unbroken vector subgroup is acceptable. The different choices simply correspond to redefinitions of the baryon field.

An anti-decuplet representation $SU(3)$ is the completely symmetric part of the tensor product of three anti-fundamentals. The components of the pentaquark anti-decuplet, with respect to the quark basis, are organized below according to the weight diagram.
\begin{align}\label{pfield}
P_{333} &= \Theta^{+}_{\bar{10}}  \\
P_{133} = \frac{1}{\sqrt{3}} \, N^0_{\bar{10}}  &\qquad 
P_{233} = \frac{1}{\sqrt{3}} \, N^+_{\bar{10}} \nn \\
P_{113} = \frac{1}{\sqrt{3}} \, \Sigma^{-}_{\bar{10}}\qquad 
P_{123} &= \frac{1}{\sqrt{6}}\Sigma^{0}_{\bar{10}}\qquad 
P_{223} = \frac{1}{\sqrt{3}} \, \Sigma^{+}_{\bar{10}}   \nn \\
P_{111} =  \Xi^{--}_{\bar{10}} \qquad 
P_{112} = \frac{1}{\sqrt{3}}  \, \Xi_{\bar{10}}^{-}  &\qquad 
P_{122} = \frac{1}{\sqrt{3}} \, \Xi_{\bar{10}}^{0} \qquad 
P_{222} =  \Xi_{\bar{10}}^{+}\, \nn
\end{align}
We will drop the $\bar{10}$ subscripts when there is no possibility for confusion. Under chiral symmetry, the field $P$ transforms as
\begin{equation}\label{ptrans}
P_{ijk}(x) \rightarrow P_{lmn}(x) U^{\dagger}(x)^l{}_i U^{\dagger}(x)^m{}_j U^{\dagger}(x)^n{}_k
\end{equation}
To construct a chiral Lagrangian for matter fields, it is convenient to trade in the $\xi$ field and its derivative for two vector fields:
\begin{align}
\label{amu} A_\mu & = \frac i 2 \left(\xi\partial_\mu \xi^\dagger - \xi^\dagger \partial_\mu \xi\right) = \frac{1}{f}\partial_\mu \pi - \frac{1}{6 f^3}\left[\left[\partial_\mu\pi, \pi \right],\pi\right] + \ldots \\
\label{vmu} V_\mu & = \frac 1 2 \left(\xi\partial_\mu \xi^\dagger + \xi^\dagger \partial_\mu \xi\right) = \frac{1}{2f^2}\left[\pi,\partial_\mu \pi \right] + \ldots
\end{align}
These vector fields have the simple transformation properties: 
\begin{align}
\label{atrans} A_\mu(x) &\rightarrow U(x) A_\mu(x) U^\dagger(x) \\ 
\label{vtrans} V_\mu(x) &\rightarrow U(x) V_\mu(x) U^\dagger(x) - \partial_\mu U U^\dagger(x).
\end{align}
In QCD, the quark mass $M_q$ explicitly breaks chiral symmetry. However, if $M_q$ is assumed to transform as
\begin{equation}
M_q \rightarrow L M_q R^\dagger
\end{equation}
then the QCD Lagrangian is invariant under chiral symmetry. In order to have the same symmetry breaking pattern in chiral perturbation theory, we have $M_q$ transform in the same way. It is also convenient to trade in $M_q$ for a combination of $M_q$ and $\xi$ which transforms like the other fields:
\begin{align}
M_q^\xi & = \left(\xi^\dagger M_q \xi^\dagger + \xi M_q^\dagger \xi\right) \\
M_q^\xi & \rightarrow U(x) M_q^\xi U^\dagger(x).
\end{align}
A covariant derivative can be defined for nucleon and pentaquark fields by
\begin{align} \label{covder}
D_\mu B &= \partial_\mu B + \left[V_\mu,B\right] \\
\left(D_\mu P\right)_{ijk} &= \partial_\mu P_{ijk} - P_{ljk} V_\mu^l{}_i - P_{ilk} V_\mu^l{}_j - P_{ijl} V_\mu^l{}_k
\end{align}
As usual, the covariant derivatives of the fields transform like the fields themselves. Under parity, the fields $B$, $V_\mu$, and $M_q^\xi$ are even, while the field $A_\mu$ is odd. Both even and odd parity will be treated for the field $P$. For most of the paper, we will assume that P is even under parity. The results for odd parity will be summarized in section \ref{negative}.

Having discussed the symmetries of the theory, we now focus on the relevant scales and the power counting in the theory. Contributions from terms in the chiral Lagrangian with derivatives acting on baryon fields are suppressed by $m_B/\Lambda_{\chi}$; however, this is not a small quantity. It is necessary to sum these contributions to all orders in $m_B/\Lambda_{\chi}$. The difficulty of doing this in practice suggests using an alternative approach called heavy baryon chiral perturbation theory (HB$\chi$PT) \cite{Jenkins:1990jv}. It uses well-tested ideas from heavy quark effective theory. In HB$\chi$PT, the momenta of baryons are split up as
\begin{equation}
p^\mu = m_B v^\mu + k^\mu,
\end{equation}
where $v \cdot k \ll m_B$ in low-energy processes. In other words, the baryons essentially have constant velocity when interacting with soft mesons. The usual baryon field $B(x)$ is traded in for baryon fields of definite velocity
\begin{equation}\label{bv}
B_v(x)=  \frac{1}{2} \left(1 + \sla{v}  \right) e^{i m_B v \cdot x} B(x),
\end{equation}
which satisfy $\sla{v} B_v = B_v$. This redefinition cancels the $m_B$ term in the leading order chiral Lagrangian resulting in a simple baryon propagator $i/(v \cdot k)$. Moreover, since the baryons are not far off-shell in low-energy processes, derivatives acting on $B_v$ produce powers of the small residual momentum $k$, which is of the size of the meson momenta, rather than the total baryon momentum $p$. Thus, a simple power counting is restored, in terms of the small quantities $k/\Lambda_{\chi}$ and $k/m_B$.

In pentaquark decays, there are two multiplets of matter fields with different masses. Applying this procedure for each field separately results in simple kinetic terms for both fields but introduces a phase in the interaction terms. An equivalent possibility is to use the same field redefinition for both fields,
\begin{equation}
P_v(x) = \frac{1}{2} \left(1 + \sla{v} \right) e^{i m_P v \cdot x} P(x) \quad \textrm{and} \quad B_v(x) = \frac{1}{2} \left(1 + \sla{v} \right) e^{i m_P v \cdot x} B(x),
\end{equation}
where $m_P$ is the mass of the pentaquarks, to keep the interaction terms simple. Doing so introduces a constant $\Delta = m_P - m_B$ in the nucleon propagator, $i/(v \cdot k + \Delta).$ We adopt this second approach, but the results are the same using either approach.

In terms of the fields $P_v$ and $B_v$, the lowest order chiral Lagrangian is $ \mathcal{ L} = \mathcal{ L}^{(1)}_P + \mathcal{L}^{(1)}_{PB} + \mathcal{ L}^{(1)}_B + \mathcal{L}^{(2)}_{\pi}$. The usual pieces involving only nucleons and mesons are,
\begin{align}
    \label{lagb} \mathcal{ L}^{(1)}_B &= \tr \left(\bar{B}_v \left(i v \cdot D + \Delta \right) B_v \right) + 2 D \tr \left(\bar{B}_v S^\mu_v \left\{A_\mu,B_v\right\}\right) + 2 F  \tr \left(\bar{B}_v S^\mu_v \left[A_\mu,B_v\right] \right) \\
    \label{lagpi} \mathcal{ L}^{(2)}_\pi &= \frac{f^2}{2} \tr \left( A^\mu A_\mu \right) + B_0 \frac {f^2}{4} \tr M_q^\xi.
\end{align}
Here $S_v^\mu$ is the (Pauli-Lubanski) spin operator in a frame with four-velocity $v$. The constants $D$ and $F$ are the usual axial vector coupling constants with $g_A = D + F$. The pieces of the Lagrangian involving pentaquarks are
\begin{align}
    \label{lagp} \mathcal{L}^{(1)}_P &= \bar{P}_v i v \cdot D P_v + 2 \mathcal{H}_{P} \bar{P}_v S_v \cdot A P_v\\
    \label{lagpb} \mathcal{L}^{(1)}_{PB} &= 2 \mathcal{C}_{P B} \left(\bar{P}_v S_v \cdot A B_v + \bar{B}_v S_v \cdot A P_v \right).
\end{align}
The constants $\mathcal{C}_{P B}$ and $\mathcal{H}_{P}$ are pentaquark analogues of constants introduced in \cite{Jenkins:1990jv} for the baryon decuplet. The contraction of flavor indices has been suppressed because there is only one possible contraction between the multiplets that results in a singlet. More explicitly, 
\begin{align}
\mathcal{L}^{(1)}_P &= \bar{P}{}^{ijk}  i v^\mu \left(D_\mu P\right)_{ijk} + 2 \mathcal{H}_{P} \bar{P}{}^{ijl} S^\mu_v A_\mu^k{}_l P_{ijk} \\
\mathcal{L}^{(1)}_{PB} &= 2 \mathcal{C}_{P B} \left(\bar{P}{}^{ijk} S^\mu_v {A_\mu}^l{}_i B^m{}_j \epsilon_{klm} + \bar{B}_m{}^j \bar{A}_\mu{\,}_l{}^i S^\mu_v P_{ijk} \epsilon^{klm} \right)
\end{align}
To be completely explicit, we should also denote the Lagrangian $\mathcal{L}$ above by $\mathcal{L}_v$. The full Lagrangian can then written as $\mathcal{L}_\textrm{full} = \int \frac{d^3 v}{2 v_0} \mathcal{L}_v$. Since we will be working in a single velocity sector, we drop this formality. Finally, note that $\mathcal{ L}^{(2)}_\pi$ would take its more familiar form in terms of the field $\Sigma = \xi^2$. 

The only constants with dimensions of mass appearing in this Lagrangian are $M_q$ and $\Delta$. In chiral perturbation theory, $M_q$ is treated as second order in the chiral expansion because $m^2_{\pi} \propto m_q$. The mass splitting in each multiplet of matter fields due to the quark masses happens at next to leading order. On the other hand, the mass difference $\Delta$ that appears in the Lagrangian does not vanish in the chiral limit; formally $\Delta \propto \Lambda_\chi$. At NLO in chiral perturbation theory, the mass difference can be expressed as 
\begin{equation}
m_P - m_B = \left(m_{\Xi_{\bar{10}}} - \sigma_P \left(\frac{2 m_K^2 + m_{\pi}^2}{\Lambda_\chi}\right)\right) - \left(m_\Sigma - \sigma_B \left(\frac{2 m_K^2 + m_{\pi}^2}{\Lambda_\chi}\right) \right)
\end{equation}
using \eq{quarkmass} below for the NLO $\Xi_{\bar{10}}$ mass and the well-known analogue for the $\Sigma$ nucleon. To obtain a rough estimate, we use the mass of the candidate $\Xi^{--}$ seen in \cite{Alt:2003vb} at $1860 \, \textrm{MeV}$ and assume that $\sigma_P$ and $\sigma_B$ are approximately equal. We find $\Delta \sim 750 \, \textrm{MeV}$. Throughout this paper, we take $\Lambda_\chi = 4 \pi f = 1.65 \, \textrm{GeV}$ so that  $\Delta / \Lambda_\chi \sim 1/2$.  This is a borderline situation that could lead to problems. Fortunately, $\Delta$ does not appear in the mass splitting up to NNLO. Thus, there are no problems in this calculation.

\begin{figure}
\caption{ \label{tree} a) A pentaquark-pion-pentaquark vertex.  b) A pentaquark-pion-nucleon vertex. This is the only diagram contributing to pentaquark decay at leading order in the chiral expansion. c) A quark mass insertion. This is the only diagram contributing to the pentaquark self-energy at NLO. Double lines are pentaquarks, single lines are nucleons, and dashed lines are pions.}
\begin{fmffile}{tree}
\fmfframe(0,0)(0,0){a)\begin{fmfgraph*}(40,20)
    \fmfdot{pbpi}
    \fmfleft{p}
    \fmfright{b}
    \fmf{double_arrow}{p,pbpi}
    \fmf{double_arrow}{pbpi,b}
    \fmffreeze
    \fmftop{pi}
    \fmf{dashes}{pbpi,pi}
\end{fmfgraph*}
\hspace{2cm}
b)\begin{fmfgraph*}(40,20)
    \fmfdot{pbpi}
    \fmfleft{p}
    \fmfright{b}
    \fmf{double_arrow}{p,pbpi}
    \fmf{plain_arrow}{pbpi,b}
    \fmffreeze
    \fmftop{pi}
    \fmf{dashes}{pbpi,pi}
\end{fmfgraph*} }
\\
\fmfframe(0,0)(0,0){c)\begin{fmfgraph*}(40,20)
    \fmfv{decor.shape=cross,decor.filled=full,decor.size=10thin}{m}
    \fmfleft{pl}
    \fmfright{pr}
    \fmf{double_arrow}{pl,m,pr}
\end{fmfgraph*} }
\end{fmffile}
\end{figure}
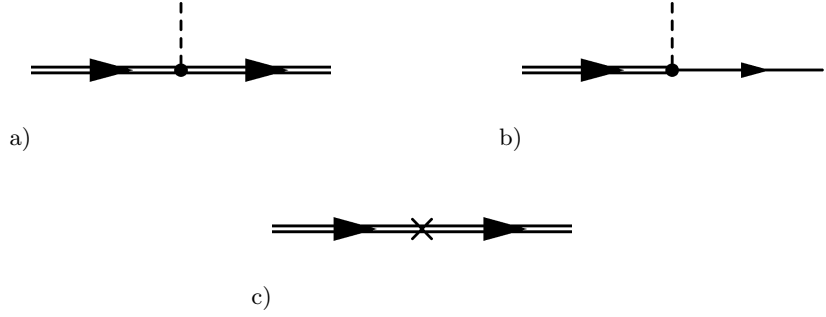

The partial widths for the decay of a pentaquark can be easily calculated from the Lagrangian at leading order. Only the tree diagram in figure \ref{tree}b
contributes at this order. The amplitude is
\begin{equation}\label{treeamp}
\mathcal{A}(P^a \rightarrow B^c \pi^e) = F^{ace} \frac{2 \mathcal{C}_{PB}}{f} \bar{B}_v k_\pi \cdot S_v P_v
\end{equation}
 where $F^{ace}$ is a flavor factor. The partial width is then
\begin{align}\label{width}
\Gamma (P^a \rightarrow B^c \pi^e) &= \frac{1}{2 m_P} \int d \Pi_f \delta(k_i - k_f) \left( \frac{1}{2} \sum_{i,f} |\mathcal{A}|^2 \right) \\ 
&= (F^{ace})^2 \frac{\mathcal{C}^2_{PB}}{2 \pi f^2} \frac{m_B}{m_P} k_\pi^3
\end{align}
In evaluating the total width of $\Theta^+$ numerically, we use the mass of the $\Theta$ for $m_P$ and the mass of the proton for $m_B$, taking into account some NLO corrections. The flavor factor $(F^{ace})^2$ for both decay channels of the $\Theta^+$ is $1$ and the momentum of the kaon produced is $269 \, \textrm{MeV}$. The total width of $\Theta^+$ is
\begin{equation}
\Gamma(\Theta^+) = \Gamma(\Theta^+ \rightarrow K^0 p) + \Gamma(\Theta^+ \rightarrow K^+ n) = \left(220 \, \textrm{MeV} \right) \, \mathcal{C}^2_{PB}.
\end{equation}
There is one minor difference between this value and that in eq.~(10) in \cite{Mehen:2004dy} due to a choice for the value of $f$ used in \eq{width}. In ordinary chiral perturbation theory, one often finds that in processes involving kaons, some of the one-loop corrections can be accounted for by replacing $f$ by $f_K$ in the tree level result. Without calculating to higher order, it is not clear that this improves the accuracy for pentaquark decays. Thus, we use the leading order $f$ for all decays.

Experiments have found the total width of $\Theta^+$ to be surprisingly small. In several experiments, the width is below the detector resolution. Two recent experiments suggest that the width is in the neighborhood of $10 \, \textrm{MeV}$ and possibly even smaller \cite{Airapetian:2003ri, Chekanov:2004kn}. Reanalysis of old high-statistics $KN$ scattering data and the non-observation of $\Theta^+$ there suggests an even smaller width, possibly smaller than $1 \, \textrm{MeV}$ \cite{Arndt:2003xz, Sibirtsev:2004bg}. The recent experimental results on the $\Theta$ width suggest that $\mathcal{C}_{PB}$ is at most $\frac{1}{4}$, and the reanalysis of old data suggests that $\mathcal{C}_{PB}$ is smaller than $\frac{1}{15}$. The small coupling constant $\mathcal{C}_{PB}$ allows a further expansion. Thus, we now have an expansion in the small quantities $k_\pi/\Lambda_\chi$, $k_\pi/m_P$, and $\mathcal{C}_{PB}$.

The small coupling argument, made for the nucleon-meson-pentaquark vertex, is also relevant for the coupling of pentaquarks to other observed baryons with masses near that of pentaquarks. Based on dynamical assumptions coming from quark models, we expect non-exotic states with a baryon and meson to have small wavefunction overlap with a pentaquark \cite{Jaffe:2003sg, Carlson:2003xb, Close:2004tp}. For example, in the diquark model, one of the diquarks in a pentaquark has to be split up in order to create a non-exotic state.

On the other hand, the pentaquark-meson-pentaquark coupling $\mathcal{H}_P$ should not be suppressed by this effect. This is in analogy with the couplings $g_A$ and $\mathcal{H}$ for the nucleons and the decuplet respectively, which have been fit to data and found to be of order $1$.

Models are needed at this stage to provide an estimate of coupling constants not measured in experiment. Chiral perturbation theory is definite about the flavor, spin, color, spatial properties of hadrons. For example, flavor Clebsch-Gordon coefficients between hadron multiplets can be calculated directly. However, chiral perturbation theory says nothing about the wavefunctions of quarks in the hadron. That is the realm of QCD. In this way, QCD determines the coupling constants appearing in the chiral Lagrangian. Although QCD itself is remarkably simple, it is complicated to compute things, such as the coupling constants.  Quark models attempt to approximate this complicated calculation. 

The small coupling argument is relevant for pentaquarks of negative parity also. The Lagrangian $\mathcal{L}_{P}$ is unchanged under a change in parity of the pentaquark field because the field appears quadratically in each term. The Lagrangian $\mathcal{L}_{PB}$, however, is affected. The chiral Lagrangian (before the heavy baryon transformations) changes by the replacement of $\gamma^\mu \gamma^5$ with $\gamma^\mu$. In the heavy baryon formulation this corresponds to replacing $2 S^\mu_v$ with $v^\mu$ between spinor fields.

This affects the determination of the constant $\mathcal{C}_{PB}$. The partial width of a negative parity pentaquark is \cite{Mehen:2004dy}
\begin{equation}\label{negwidth}
\Gamma_{-}(P^a \rightarrow B^c \pi^e)= (F^{ace})^2 \frac{\mathcal{C}_{-}^2{}_{PB}}{2 \pi f^2} \frac{m_B}{m_P} E_\pi^2 k_\pi.
\end{equation}
The width for negative parity differs from the width for positive parity only in the replacement of $k_\pi^3$ by $E_\pi^2 k_\pi$. For the total width of the $\Theta$, we find
\begin{equation}\label{negwidthnum}
\Gamma_{-}(\Theta)= (966 \, \textrm{MeV}) \, \mathcal{C}_{-}^2{}_{PB}.
\end{equation}
Thus, for negative parity, recent experiments suggest that $\mathcal{C}_{-}{}_{PB}$ is at most $\frac{1}{8}$, and the reanalysis of old $KN$ scattering data suggests that $\mathcal{C}_{-}{}_{PB}$ is smaller than $\frac{1}{30}$. In other words, the small coupling expansion is even more reliable for a pentaquark of negative parity. Of course, we could look at this another way. It would be surprising for a pentaquark of very small width to have negative parity, as it requires an extremely small coupling and a corresponding dynamical mechanism. 

At NLO, the quark mass enters the pentaquark Lagrangian:
\begin{equation}\label{quarkmass}
 \mathcal{L}_P^{(2)} = b_\mathcal{H} \frac{B_0}{4 \pi f} \bar{P}_v M_q^\xi P_v
+ \sigma_P \frac{B_0}{4 \pi f} (\tr M_q^\xi) \bar{P}_v P_v + \ldots 
\end{equation}
Here $B_0$ is the constant appearing in the meson Lagrangian in \eq{lagpi}. The factor of $\frac{B_0}{4 \pi f}$ has been included so that NLO mass shifts are in terms of an order one coupling and $\frac{m_\pi^2}{4 \pi f}$, rather than $m_q$. An alternate definition of the coupling constants $b_\mathcal{H}$ and $\sigma_P$ would be to leave out this factor. The dots in \eq{quarkmass} refer to terms that do not contribute to the self-energy at NLO.

The quark mass contributes to the pentaquark self-energy at NLO as shown in figure \ref{tree}b. We work in the isospin limit, with $m_u = m_d = \hat m$. Isospin breaking, by quark mass differences and by electromagnetism, is negligible numerically, until a higher order in chiral perturbation theory. In the isospin limit, the quark masses are related to the squared pion masses by
\begin{equation} \label{pionmass}
m_\eta^2 =  B_0 (\frac{4}{3} m_s + \frac{2}{3} \hat m), \quad  m_K^2 = B_0 (m_s + \hat m), \quad m_\pi^2 = 2 B_0 \hat m.
\end{equation}
For the pentaquark self-energy diagram in figure \ref{tree}c, we find 
\begin{equation}\label{NLOself}
\Sigma^{(2)}_a = \alpha^a_K \frac{m^2_K}{4 \pi f} + \alpha^a_\pi \frac{m^2_\pi}{4 \pi f}.
\end{equation}
where the index $a$ runs over pentaquarks. Summing powers of the self-energy in the usual way shifts the mass in the propagator, $m_a = m_P + \Sigma^{(2)}_a$. The constants $\alpha^a_K$ and $\alpha^a_\pi$ in \eq{NLOself} are 
\begin{align} \label{NLOshift}
\alpha^\Theta_K &=  - 2 b_\mathcal{H}  -2 \sigma_P & \alpha^\Theta_\pi &=   b_\mathcal{H}  - \sigma_P\\
\nn \alpha^N_K &= - \frac{4}{3} b_\mathcal{H}  -2 \sigma_P & \alpha^N_\pi &=  \frac{1}{3} b_\mathcal{H}  - \sigma_P \\
\nn \alpha^\Sigma_K &= - \frac{2}{3} b_\mathcal{H} -2 \sigma_P & \alpha^\Sigma_\pi &= - \frac{1}{3} b_\mathcal{H} - \sigma_P \\
\nn \alpha^\Xi_K &=   -2 \sigma_P & \alpha^\Xi_\pi &= - b_\mathcal{H} - \sigma_P.
\end{align}
There are four isospin multiplet masses and two parameters. Thus, there are two mass relations analogous to Gell-Mann's equal spacing rule for isospin multiplets in the decuplet,
\begin{equation}\label{esr}
m_\Xi - m_\Sigma = m_\Sigma - m_N = m_N - m_\Theta.
\end{equation}
Using the mass of the candidate $\Xi^{--}$ seen in \cite{Alt:2003vb} at $1860 \, \textrm{MeV}$ and the relation
\begin{equation}\label{bh}
 m_\Xi - m_\Theta =  b_\mathcal{H} \left(\frac{m_K^2 - m_\pi^2}{2 \pi f}\right),
\end{equation}
we find $b_\mathcal{H} = 1.17$. Models for the pentaquark anti-decuplet \cite{Jaffe:2003sg,Diakonov:1997mm} suggest that the constant $b_\mathcal{H}$ of order $1$. The $\sigma_P$ term in the Lagrangian only contributes to an overall shift to the masses at this order; however, it cannot be eliminated entirely since $M_q^\xi$ also contributes vertices with pions.

The mass relations obtained here differ considerably from those in \cite{Jaffe:2003sg} because we do not consider mixing between the anti-decuplet and a degenerate octet. In particular, without mixing, the $\Theta$ is the lightest member of the anti-decuplet at $1540 \, \textrm{MeV}$. The Roper resonance at $1440 \, \textrm{MeV}$ cannot be accommodated in the anti-decuplet. We will take up the inclusion of a degenerate octet in chiral perturbation theory elsewhere.

\section{Power Counting} \label{count}
To calculate to higher orders consistently, we need to establish the power counting for this theory. The coupling $\mathcal{C}_{PB}$ was estimated to be $\frac{1}{4}$ or smaller. Since it is about the same size as $\frac{m_K}{4 \pi f}$, we will expand jointly in the chiral order and the order in the coupling constant. The order of a generic diagram is
\begin{equation}
D = 4 L - 2I_\pi - I_H  + \sum_k \; k \left( N^k_{\pi} + N^k_{H} \right) + N_{\mathcal{C}}.
\end{equation}
The last term in this equation simply accounts for each of the $N_{\mathcal{C}}$ instances of a small coupling. It includes the small pentaquark-pion-nucleon coupling $\mathcal{C}_{PB}$ and possibly other couplings expected to be small by our dynamical assumption, such as $b_{\mathcal{C}}$ in \eq{PMqB}. The remaining terms in the equation account for the chiral order of the diagram. With a mass-independent renormalization scheme, the only dimensional quantities appearing are the small momenta. Thus, it is sufficient to determine the factors of momentum appearing in the diagram. There is a factor of $4$ for each loop from $d^4p$, a factor of $-2$ for each internal meson line from the propagator, and a factor of $-1$ for each internal heavy particle line from the propagator.  There is a factor of $k$ for each of the $N^k_\pi$ vertices from the $k^\textrm{th}$ order meson Lagrangian and for each of the $N^k_H$ vertices from the $k^\textrm{th}$ order heavy particle Lagrangian. These terms account for the chiral order of the diagram.

In addition, every diagram satisfies the identities, 
\begin{align}\label{graphids}
  \left( I_\pi + I_H \right) - \left( \sum_k \; N^k_{H} + N^k_{\pi} \right)+ L = 1 \\
  2 I_H + E_H = \sum_k \;  2 N^k_{H}. 
\end{align}  
Here $E_H$ is the number of external heavy particle lines. The first of these equations is an analogue of Euler's formula for a Feynman diagram; the second equation simply counts the ``ends" of heavy particle lines in two different ways. Combining these three equations yields the power counting rule,
\begin{equation}\label{power}
D= 2 L + 2 - \frac{1}{2}E_H + \sum_k \left[ (k-1) N^k_H + (k-2) N^k_\pi \right] + N_{\mathcal{C}}. 
\end{equation}
If all the heavy particles are nucleons and there are no instances of the small coupling, this reduces to the power counting rule for nucleon chiral perturbation theory. Since each term in \eq{power} is nonnegative, the diagrams contributing at a given order can be enumerated systematically. 

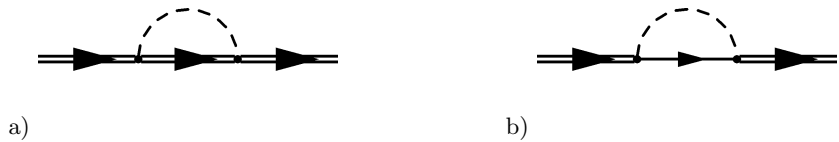
\begin{figure}[b]
\caption{\label{loop} Self-energy corrections appear at third order in the chiral expansion. Double lines are pentaquarks, single lines are nucleons, and dashed lines are pions.}
\vspace{.5cm}
\begin{fmffile}{se}
\fmfframe(1,1)(1,1){a) \begin{fmfgraph*}(40,20)
    \fmfleft{p1}
    \fmfright{p2}
    \fmfv{decor.shape=circle,decor.filled=full,decor.size=2thin}{pi1}
    \fmfv{decor.shape=circle,decor.filled=full,decor.size=2thin}{pi2}
    \fmf{double_arrow,straight}{p1,pi1,pi2,p2}
    \fmf{dashes,left,tension=.01}{pi1,pi2}
\end{fmfgraph*}
\hspace{2cm}
b) \begin{fmfgraph*}(40,20)
    \fmfleft{p1}
    \fmfright{p2}
    \fmfv{decor.shape=circle,decor.filled=full,decor.size=2thin}{pi1}
    \fmfv{decor.shape=circle,decor.filled=full,decor.size=2thin}{pi2}
    \fmf{phantom, straight}{p1,pi1,pi2,p2}
    \fmf{double_arrow}{p1,pi1}
    \fmf{double_arrow}{pi2,p2}
    \fmf{plain_arrow}{pi1,pi2}
    \fmf{dashes,left,tension=.01}{pi1,pi2}
\end{fmfgraph*} }
\end{fmffile}
\end{figure}

For example, the quark mass insertion in figure \ref{tree}c contributes to the mass splitting of pentaquarks at second order as already discussed in \eq{NLOself}. At third order, loop diagrams arise; however, they can only contain vertices from the first order Lagrangian in \eq{lagp}. There are two such diagrams with third {\it chiral} order, as in figure \ref{loop}. The nucleon on the internal line in the second diagram suppresses it by two factors of the small coupling $\mathcal{C}_{PB}$ relative to the first. The calculation of the diagrams is taken up in the next section.

\section{Masses at NNLO} \label{mass}

The power counting developed in the previous section determines what diagrams could contribute to the mass shift at NNLO. There are one-loop diagrams with vertices from $\mathcal{L}^{(1)}$, tree diagrams with vertices from $\mathcal{L}^{(1)}$ and at most two vertices from $\mathcal{L}^{(2)}$, and tree diagrams with vertices from $\mathcal{L}^{(1)}$ and at most one vertex from $\mathcal{L}^{(3)}$. Fortunately, the actual number of diagrams that contribute is small. Vertices in $\mathcal{L}^{(2)}$ and $\mathcal{L}^{(3)}$ that involve derivatives acting on pentaquarks fields contribute to the kinetic energy but do not affect the mass shift, i.e. the self-energy at zero momentum. In other words, the relevant terms in the higher order Lagrangians involve only the quark mass. Since $M_q$ is second order, terms with factors of $M_q$ and no derivatives appear only in the even order Lagrangians. The relevant terms in $\mathcal{L}^{(2)}_{P}$ appear in \eq{quarkmass}, and the relevant terms in $\mathcal{L}^{(2)}_{PB}$ are
\begin{equation}\label{PMqB}
\mathcal{L}^{(2)}_{PB} = b_\mathcal{C} \left(\bar{P}_v M_q^\xi B_v + \bar{B}_v M_q^\xi P_v \right) + \ldots .
\end{equation}

\begin{figure}
  \caption{\label{xx} Two insertions of the quark mass term in $\mathcal{L}^{(2)}_{PB}$}
\begin{fmffile}{xx}
\fmfframe(1,1)(1,1){\begin{fmfgraph*}(40,20)
    \fmfv{decor.shape=cross,decor.filled=full,decor.size=10thin}{m1}
    \fmfv{decor.shape=cross,decor.filled=full,decor.size=10thin}{m2}
    \fmfleft{pl}
    \fmfright{pr}
    \fmf{double_arrow}{pl,m1}
    \fmf{double_arrow}{m2,pr}
    \fmf{plain_arrow}{m1,m2}
\end{fmfgraph*} }
\end{fmffile}
\end{figure}
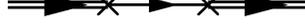

It would seem then that there are two diagrams contributing to the mass shift at NNLO. There is the diagram in figure \ref{xx} and the one-loop diagram in figure \ref{loop}a. (The analogue of figure \ref{xx} with the internal baryon line replaced by a pentaquark is not one-particle-irreducible; its contribution is already included in the NLO mass shift.) The first of these is proportional to $b_\mathcal{C}^2$. Note the similarity of the $b_\mathcal{C}$ term in \eq{PMqB} with the $\mathcal{C}_{PB}$ term in \eq{lagpb}. In fact, quark models suggest that the constant $b_\mathcal{C}$ should be small like $\mathcal{C}_{PB}$. Since the diagram in figure \ref{xx} is further suppressed by this small constant, it is higher order. That leaves one one-loop diagram contributing to the mass shift at NNLO. 

The diagram in figure \ref{loop}a evaluates to
\begin{equation}\label{sep}
- i \Sigma^{(3)}_a(k) = \frac{4 \mathcal{H}_P^2}{f^2} \bar{P}_v S_v^\mu S_v^\lambda P_v \sum_f G_{a f} K^f_{\mu \lambda} (-v \cdot k)
\end{equation}
where $a$ and $b$ label pentaquarks and $f$ is summed over mesons. The object $P_v$ in this equation is just a spinor, while $G_{a f}$ has all the flavor dependence,
\begin{equation} \label{flav}
G_{a f} = ({\bar{P}{}^a{\,}^{i j k}} \pi^f{}_k{}^l P^g{\,}_{i j l})({\bar{P}{}^g{}^{m n p}} \pi^f{}^q{}_m P^a{}_{n p q}),
\end{equation}
where all repeated indices other than $a$ and $f$ are summed over. The function $K$ comes from the loop integral,
\begin{equation}
K_{\mu \lambda}^f(\omega) = \int \frac{d^4l}{(2 \pi)^4} \frac{l_\mu l_\lambda}{(v \cdot l - \omega) (l^2 - m_f^2)}.
\end{equation}
Calculating the loop integrals at zero external momentum, we find
\begin{equation}\label{sigma0}
- i \Sigma^{(3)}_a(0) = \frac{4 \mathcal{H}_P^2}{f^2} \bar{P}_v S^2 P_v \sum_f G_f{}^{a a} \left(\frac{-i}{24 \pi} m_f^3 \right).
\end{equation}
Thus, up to NNLO the mass shift is
\begin{equation} \label{mshift}
m_a = m_P + \alpha^a_K (\frac{m_K}{4 \pi f}) m_K + \alpha^a_\pi (\frac{m_\pi}{4 \pi f})m_\pi - \beta^a_\eta (\frac{m_\eta}{4 \pi f})^2 m_\eta - \beta^a_K (\frac{m_K}{4 \pi f})^2 m_K - \beta^a_\pi (\frac{m_\pi}{4 \pi f})^2 m_\pi
\end{equation}
where $\alpha^a_f$ are the NLO shifts in \eq{NLOshift} and $\beta^a_f = 2 \pi \mathcal {H}_P^2 G_f{}^{a a} $. The similarity between the forms of the NLO and NNLO terms is a bit misleading. The NLO terms are often expressed in terms of the quark mass. We have adopted a convention for the quark mass term in the pentaquark Lagrangian that allows us to use the LO relations for the squared pion masses in \eq{pionmass} to write the NLO terms in \eq{mshift} in terms of the pion masses. The pion masses appearing in the NNLO terms on the other hand are physical, coming from pion loops. The same is true for the factors of $4 \pi f$; those in the NLO terms are conventional while those in the NNLO terms are physical, coming from pion loops. Of course, it's possible to turn this around and write both NLO and NNLO terms in terms the quark masses. As an example, we do this for the $\Theta$ mass, 
\begin{equation} \label{thetamass}
m_\Theta = m_P - 2(\tilde b_{\mathcal{H}}+ \tilde \sigma_P) m_s + 2 (\tilde b_{\mathcal{H}} - 2 \tilde \sigma_P)\hat m - \frac{ {\mathcal{H}}_P^2 B_0^{\frac{3}{2}}}{12 \pi f^2} \left[(\frac{4}{3}m_s + \frac{2}{3} \hat m)^{\frac{3}{2}} + (m_s + \hat m)^{\frac{3}{2}}\right],
\end{equation}
where the tildes reflect an alternative definition of the coupling constants. The calculation of flavor factors such as in \eq{flav} for loop diagrams is greatly simplified using Mathematica. The result for $\beta^a_f$ is
\begin{align}\label{beta}
\beta^\Theta_\eta &=   \frac{4 \pi}{3} \mathcal{H}_P^2 & \beta^\Theta_K &=   \frac{4 \pi}{3} \mathcal{H}_P^2 & \beta^\Theta_\pi &=  0 \\
\nn \beta^N_\eta &=  \frac{\pi}{3} \mathcal{H}_P^2 & \beta^N_K &= 2 \pi \mathcal{H}_P^2 & \beta^N_\pi &=  \frac{\pi}{3} \mathcal{H}_P^2\\
\nn \beta^\Sigma_\eta &=  0 & \beta^\Sigma_K &= \frac{16 \pi}{9} \mathcal{H}_P^2 & \beta^\Sigma_\pi &=  \frac{8 \pi}{9} \mathcal{H}_P^2 \\
\nn \beta^\Xi_\eta &=  \frac{\pi}{3} \mathcal{H}_P^2 & \beta^\Xi_K &= \frac{2 \pi}{3} \mathcal{H}_P^2 & \beta^\Xi_\pi &=  \frac{5 \pi}{3} \mathcal{H}_P^2.
\end{align}
In the limit of massless up and down quarks, the numbers in \eq{beta} reduce to an inverted form of the corresponding numbers for the decuplet \cite{Jenkins:1991ts} up to a constant factor coming from the spins.

To summarize, at leading order the pentaquark masses are equal. The average mass of the anti-decuplet is $m_P$ and the spacing between isospin multiplets is $0$. At NLO, the $b_{\mathcal{H}}$ quark mass term in the Lagrangian produces an equal spacing between isospin multiplets. The mass difference $m_N - m_\Theta$ and its two analogues are nonzero and equal to $\delta^{NLO}$ where
\begin{equation}\label{spacingNLO}
    \delta^{NLO} = b_{\mathcal{H}}\left(\frac{m_K^2 - m_{\pi}^2}{6 \pi f}\right).
\end{equation}
A measurement of the spacing and \eq{spacingNLO} determines the constant $b_{\mathcal{H}}$ at NLO. As mentioned earlier, using the mass of the candidate $\Xi^{--}(1860)$ seen in \cite{Alt:2003vb}, we find $b_{\mathcal{H}} = -1.17$ at NLO. In addition, both the $b_{\mathcal{H}}$ term and the $\sigma_P$ term shift the average mass of the anti-decuplet to
\begin{equation}\label{aveNLO}
    m^{NLO}_{ave} = m_P - (b_{\mathcal{H}}+ 3 \sigma_P)\left( \frac{2 m_K^2 + m_{\pi}^2}{12 \pi f}\right).
\end{equation}
The data on masses are insufficient to determine $m_P$ and $\sigma_P$ independently. Using \eq{NLOshift}, we find
\begin{equation}\label{paramNLO}
    m_P - \sigma_P \left( \frac{2 m_K^2 + m_{\pi}^2}{4 \pi f}\right) = 1860 \, \textrm{MeV}
\end{equation}

At NNLO, the pion-induced self-energy shifts the masses. There is a shift in the average mass, a contribution to the {\it average} spacing between isospin multiplets, and a correction to equal spacing. The average mass of the anti-decuplet at NNLO is
\begin{equation}\label{aveNNLO}
    m^{NNLO}_{ave} = m^{NLO}_{ave} - \mathcal{H}_P^2 \left( \frac{m_{\eta}^3 + 4 m_K^3 + 3 m_{\pi}^3}{48 \pi f^2}\right).
\end{equation}
Although the NNLO shift is suppressed by one chiral order or about $\frac{1}{3}$, it turns out that the self-energy integral contributes an extra $2 \pi$. Thus, the NNLO shift in the average mass could be comparable to the NLO shift. This simply means that the NNLO fit for the combination of $m_P$ and $\sigma_P$ in \eq{paramNLO} could differ from the NLO fit more than expected.

The average spacing between isospin multiplets at NNLO is
\begin{equation}\label{spacingNNLO}
    \delta^{NNLO} = \delta^{NLO} + \mathcal{H}_P^2 \left( \frac{3 m_{\eta}^3 + 2 m_K^3 - 5 m_{\pi}^3}{144 \pi f^2}\right)
\end{equation}
The NNLO contribution to the average spacing could also be comparable to the NLO contribution. Thus, the NNLO fit for $b_{\mathcal{H}}$ could differ from the NLO fit more than expected.

At NNLO, the difference $(m_\Sigma- m_N) - (m_N - m_\Theta)$ and its two analogues are nonzero and equal to $\epsilon^{NNLO}$ where  
\begin{align}\label{corrNNLO}
\epsilon^{NNLO} &= - \mathcal{H}_P^2 \left( \frac{3 m_{\eta}^3 - 4 m_K^3 + m_{\pi}^3}{72 \pi f^2}\right) \\
&= \mathcal{H}_P^2 \left( 1.31 \, \textrm{MeV} \right).
\end{align}
Unlike the NNLO effects in \eq{aveNNLO} and \eq{spacingNNLO}, $\epsilon^{NNLO}$ is suppressed significantly relative to NLO effects like $\delta^{NLO}$. In principle, a measurement of $\epsilon^{NNLO}$ determines $\mathcal{H}_P$, which can then be used to determine $b_{\mathcal{H}}$ at NNLO via \eq{spacingNNLO}. Current data are insufficient to carry through this analysis. As mentioned earlier, based on quark models we expect that $\mathcal{H}_P$ is of order $1$. In this case, the correction to equal spacing is roughly $1 \, \textrm{MeV}$. In other words, we expect the corrections to the equal spacing rule to be negligible and difficult to measure experimentally. It is possible that such a cancellation does not occur at NNNLO. Dimensional analysis and a renormalization scale of $\mu = 4 \pi f$ suggest corrections of about $15 \, \textrm{MeV}$.

Thus, the problem of determining $\mathcal{H}_P$, and of determining $b_{\mathcal{H}}$ at NNLO, remains. In addition, unlike $\mathcal{C}_{PB}$, $\mathcal{H}_P$ cannot be determined using decays because decays within the anti-decuplet are kinematically forbidden. One possibility that needs to be explored further is the determination of $\mathcal{H}_P$ and $\tilde b_\mathcal{H} +  2 \tilde \sigma_P$ on the lattice using the behavior of the pentaquark mass as a function of the light quark mass in \eq{thetamass}. Lattice work on pentaquarks is well underway. For example, Csikor et. al. identify a negative parity candidate for $\Theta^{+}$ \cite{Csikor:2003ng} .

The large but benign NNLO effects described above were originally studied in the nucleon and decuplet sector in \cite{Jenkins:1991ts}. In that paper, it was also pointed out that, in the limit of massless up and down quarks, the difference $\epsilon^{NNLO}$ contains a suppression factor of $(\frac{2}{\sqrt 3} - 1)$. For example, 
\begin{equation}\label{corrNNLOJenkins}
\epsilon^{NNLO} = \mathcal{H}_P^2 \left(\frac{2}{\sqrt 3} - 1\right) \frac{m_K^3}{18 \pi f^2},
\end{equation}
which is about $20 \, \textrm{MeV}$ for $\mathcal{H}_P$ of order 1. In \cite{Jenkins:1995gc}, flavor $27$ contributions to the mass, such as those coming from pion loops, were studied in greater generality. It was noted that the suppression of flavor $27$ contributions is even greater with nonzero up and down quark masses, as seen in \eq{corrNNLO} above.

At one higher order, there are contributions from one-loop diagrams with quark mass insertions on the internal line, from wavefunction renormalization of the external lines in the tree level diagram, and from terms in the fourth order Lagrangian with two factors of the quark mass. The first two of these contribute logarithmic terms like $\frac{m_K^4}{(4 \pi f)^3} \ln{(\frac{m_K^2}{\mu^2})}$ to the mass shift. One-loop diagrams and quark mass insertion diagrams with two factors of a small coupling such as in figure \ref{loop}b and figure \ref{xx} appear at one higher order yet.

\section{Negative parity} \label{negative}

The calculation of the mass shift can be easily repeated for pentaquarks of negative parity. As discussed earlier, the Lagrangian $\mathcal{L}_{P}$ is unchanged under a change in parity of the pentaquark field because the field appears quadratically in each term. The Lagrangian $\mathcal{L}_{PB}$ is altered, but this affects only the determination of the constant $\mathcal{C}_{PB}$.

The mass shift calculation proceeds exactly as in the positive parity case. Only the self-energy diagrams with pentaquarks coupling to non-exotic baryons, in figures \ref{loop}b and \ref{xx} are affected. As pointed out in the previous section, these diagrams are fifth order. Thus, the mass shifts are not sensitive to parity even at NNNLO, which includes leading logarithms.

This suggests that negative and positive parity states should behave similarly under chiral extrapolation. In particular, the ordering between these states should remain unchanged under chiral extrapolation. This result holds even if a pentaquark octet is included, as long as it is of the same parity as the anti-decuplet.

\section{Conclusion}
Heavy baryon chiral perturbation provides a framework to calculate the mass splitting in the pentaquark anti-decuplet systematically up to NNLO. The calculation could be complicated by baryons in the nearby energy range. However, an expansion in the coupling of pentaquarks to non-exotic baryons simplifies calculations. With this coupling expansion it turns out that the mass splitting is independent of parity even at NNNLO. Corrections to the equal spacing rule at NNLO are negligible. Our results for the pentaquark mass splitting as a function of the pion masses should be helpful in chiral extrapolation of lattice results. In addition, lattice methods could help determine the two coupling constants in the chiral Lagrangian. In future work, it will be important to consider the effects of the other pentaquark multiplets predicted by spin-flavor symmetry in diquark models, uncorrelated quark models, or large $N$.

\subsection{Acknowledgements}
I would like to thank Iain Stewart for many helpful conversations. I would also like to thank Bob Jaffe, Aneesh Manohar, and Elizabeth Jenkins for comments. This work was supported in part by the NSF under the Mazur/Taubes research grant and by the DOE under cooperative research agreement DF-FC02-94ER40818.

\end{document}